\documentclass[conference]{IEEEtran}
\IEEEoverridecommandlockouts

\usepackage{cite}
\usepackage{hyperref}
\usepackage{amsmath,amssymb,amsfonts}
\usepackage{algorithmic}
\usepackage{graphicx}
\graphicspath{{./figure/}}
\usepackage{textcomp}
\usepackage{xcolor}
\usepackage{multirow, utfsym}

\usepackage[a4paper, total={184mm,239mm}]{geometry}
\def\BibTeX{{\rm B\kern-.05em{\sc i\kern-.025em b}\kern-.08em
    T\kern-.1667em\lower.7ex\hbox{E}\kern-.125emX}}

\DeclareRobustCommand*{\IEEEauthorrefmark}[1]{%
  \raisebox{0pt}[0pt][0pt]{\textsuperscript{\footnotesize #1}}%
}

\begin{document}

\title{Column-wise Quantization of Weights\\ and Partial Sums for Accurate and Efficient\\ Compute-In-Memory Accelerators
\thanks{\textsuperscript{*}Corresponding authors.}
}

\author
{\IEEEauthorblockN
{Jiyoon Kim\IEEEauthorrefmark{1},
Kang Eun Jeon\IEEEauthorrefmark{2},
Yulhwa Kim\IEEEauthorrefmark{3}\textsuperscript{*}, and
Jong Hwan Ko\IEEEauthorrefmark{2}\textsuperscript{*}}
\IEEEauthorblockA{\IEEEauthorrefmark{1}Department of Artificial Intelligence, \IEEEauthorrefmark{2}Department of Electrical and Computer Engineering}
\IEEEauthorblockA{\IEEEauthorrefmark{3}Department of Semiconductor Systems Engineering,
Sungkyunkwan University}
\IEEEauthorblockA{\{jiyun19, kejeon, yulhwakim, jhko\}@skku.edu}
}

\maketitle

\begin{abstract}
Compute-in-memory (CIM) is an efficient method for implementing deep neural networks (DNNs) but suffers from substantial overhead from analog-to-digital converters (ADCs), especially as ADC precision increases.
Low-precision ADCs can reduce this overhead but introduce partial-sum quantization errors degrading accuracy.
Additionally, low-bit weight constraints, imposed by cell limitations and the need for multiple cells for higher-bit weights, present further challenges.
While fine-grained partial-sum quantization has been studied to lower ADC resolution effectively, weight granularity, which limits overall partial-sum quantized accuracy, remains underexplored.
This work addresses these challenges by aligning  weight and partial-sum quantization granularities at the column-wise level.
Our method improves accuracy while maintaining dequantization overhead, simplifies training by removing two-stage processes, and ensures robustness to memory cell variations via independent column-wise scale factors.
We also propose an open-source CIM-oriented convolution framework to handle fine-grained weights and partial-sums efficiently, incorporating a novel tiling method and group convolution.
Experimental results on ResNet-20 (CIFAR-10, CIFAR-100) and ResNet-18 (ImageNet) show accuracy improvements of 0.99\%, 2.69\%, and 1.01\%, respectively, compared to the best-performing related works.
Additionally, variation analysis reveals the robustness of our method against memory cell variations.
These findings highlight the effectiveness of our quantization scheme in enhancing accuracy and robustness while maintaining hardware efficiency in CIM-based DNN implementations.
Our code is available at \url{https://github.com/jiyoonkm/ColumnQuant}.
\end{abstract}


\section{Introduction}
In recent years, compute-in-memory (CIM) has become an efficient paradigm for implementing deep neural networks (DNNs)\cite{b1, b2}, reducing data transfer between memory and computational units.
Nevertheless, CIM-based architectures face significant analog-to-digital converter (ADC) overhead, which increases with ADC precision, affecting both area and energy consumption\cite{b1, b2}.
Low-resolution ADCs alleviate this issue but require partial-sum quantization, introduces errors that degrade network accuracy.
Furthermore, accuracy is further degraded by low-bit weight constraints due to cell representation limits and the need for multiple cells to support higher-bit weights.
Consequently, the efficiency and accuracy also heavily rely on weight quantization\cite{b12} in CIM-based DNNs.

To effectively lower the ADC resolution, the partial-sum quantization has been explored in previous works\cite{b5, b6, b7, b8, b9}.
Finer granularity in partial-sum quantization, such as array-wise\cite{b6,b7,b8} and column-wise\cite{b9}, has played a key role in enhancing the accuracy of DNN models with low-bit partial-sums.
Partial-sum quantization has also progressed from a post-training quantization (PTQ) approach\cite{b5,b6,b7} to a quantization-aware training (QAT)\cite{b8,b9}, enabling models to recover accuracy loss during training.
As a result, previous works have restored the accuracy of partial-sum quantized models to levels comparable to those without partial-sum quantization.

Meanwhile, the weight quantization granularity is crucial since low-bit weight models set an upper bound for the accuracy achievable with partial-sum quantization.
Although some attempts have refined weight granularity\cite{b6,b7}, the improvements have been modest and insufficient to fully optimize performance, leaving the potential for further enhancement.
\begin{figure}[tbp]
    \centering
    \includegraphics[width=\columnwidth]{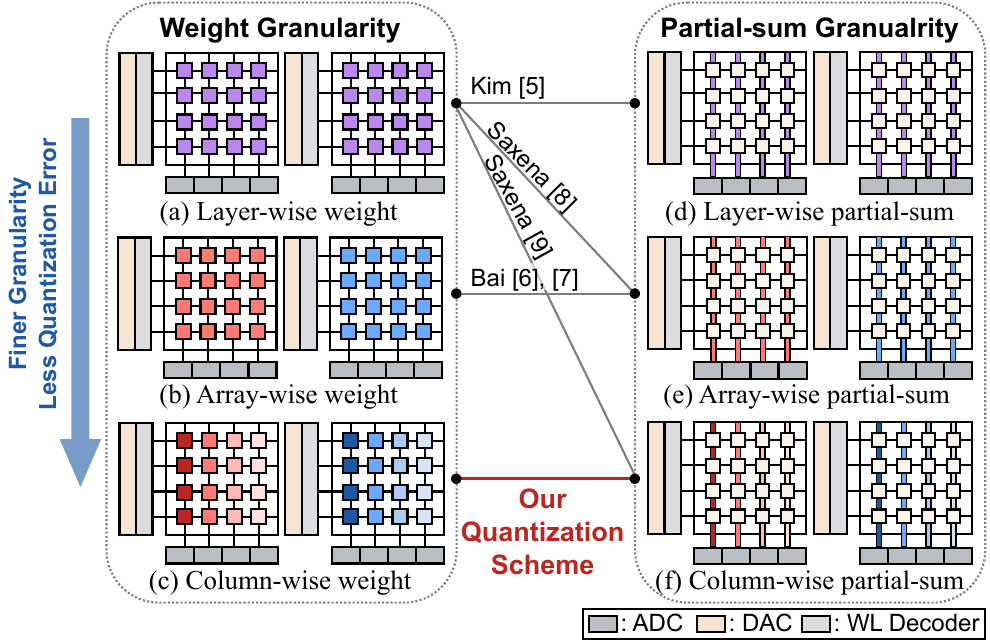}
    \vspace*{-8mm}
    \caption{Overview of the proposed quantization method and previous works.}
    \label{fig:intro}
    \vspace*{-7mm}
\end{figure}

To overcome these obstacles, we introduce an innovative strategy that aligns weight and partial-sum quantization granularity, specifically at the column-wise level as illustrated in Fig.~\ref{fig:intro}.
Our method offers several key advantages: first, it provides finer control over quantization, allowing column-wise weight quantization to capture weights accurately, which in turn enables more precise column-wise partial-sum quantization.
This improved accuracy is achieved without increasing dequantization overhead.
In addition, it enhances training efficiency by eliminating the need for two-stage training, typically required when the granularities differ.
Finally, it is robust to memory cell variations due to independent column-wise scale factors. 

However, implementing column-wise quantization poses challenges due to the time costs and increased complexity associated with fine-grained weights and partial-sums.
Mitigating these difficulties, we propose the first open-source framework designed for CIM-oriented convolution with a novel array tiling method that preserves stretched kernels within each array, removing the bottlenecks of the im2col approach.
Group convolution further eliminates sequential convolution delays and simplifies access to array-wise partial-sums.

Our experiments on ResNet-20 with CIFAR-10 and CIFAR-100, and ResNet-18 with ImageNet show notable accuracy improvements of 0.99\%, 2.69\%, and 1.01\%, respectively compared to the best-performing related works.
Furthermore, variation analysis confirms the robustness of our method to memory cell variations, emphasizing its effectiveness in improving both accuracy and hardware efficiency in CIM-based DNNs.

\section{Background \& Related Work}

\subsection{DNN Inference with CIM}
\begin{figure}[tbp]
    \centerline{\includegraphics[width=\columnwidth]{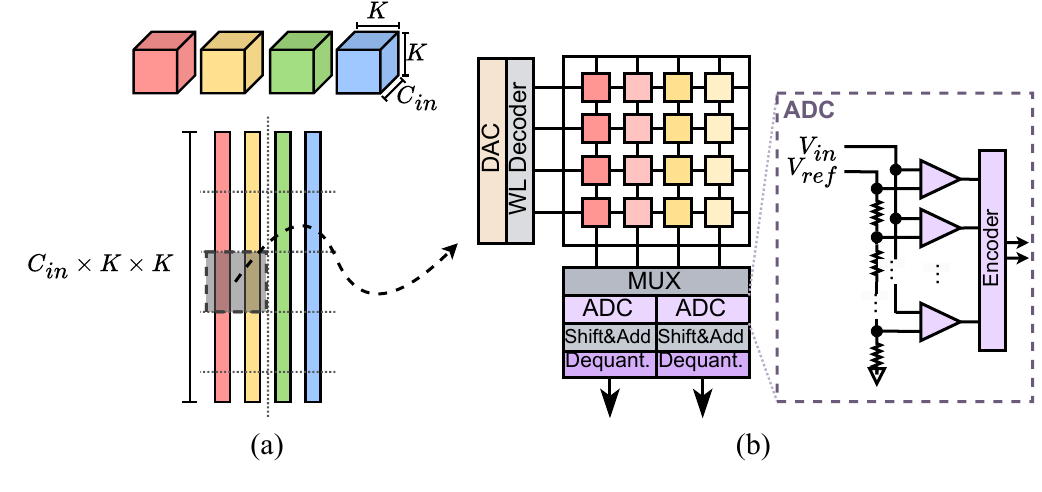}}
    \vspace*{-5mm}
    \caption{
    Implementation of the convolution layer on bit-scalable CIM architecture.
    (a) Im2col mapping and tiling process.
    $C_{in}$ represents the number of input channels, and $K$ denotes the kernel size.
    (b) Bit-scalable CIM architecture.}
    \label{fig:cim}
    \vspace*{-6.5mm}
\end{figure}

Fig.~\ref{fig:cim} briefly illustrates an overview of CIM mapping, tiling, and architecture, where the bit-scalable multiplication-accumulation(MAC) operations are performed directly within the memory array\cite{b1, b2}.
In Fig.~\ref{fig:cim}(a), convolutional weights are mapped into CIM arrays and then tiled to match the size of the array.
In this example, image-to-column (im2col) mapping is used, where each convolutional kernel is stretched into a column vector.
In Fig.~\ref{fig:cim}(b), weights are stored across multiple cells depending on the weight bit precision and the number of bits per cell.
Inputs, provided through the word lines(WLs), are processed via a digital-to-analog converter (DAC) and fed into the memory cells.
Within the array, weights and input bits are multiplied, generating partial-sums as electrical currents.
After the MAC operation, partial-sums are routed through a multiplexer, and digitized by ADCs introducing severe overhead in terms of area and energy consumption.
The reference voltage for each ADC, $V_{ref}$, is set by the scale factor corresponding to its input partial-sums, ensuring that the digitization process accurately captures their varying magnitudes.
Once digitized, the partial-sums undergo a shift-and-add operation.
Finally, the partial-sums are dequantized, with the stored scale factors multiplied to restore the original values as closely as possible.

\subsection{Related Work}
Several studies have proposed solutions for the ADC challenge in CIM, but major differences remain in the quantization scheme, training strategy, and learnable scale factor utilization.
Specifically, the quantization scheme refers to the quantization granularity where elements are grouped and assigned a common scale factor during quantization.
Finer granularities apply unique scale factors to smaller groups, while coarser granularities use a single scale factor for larger groups.
In CIM-based architectures, quantization granularities for both weights and partial-sums range from conventional layer-wise to column-wise, as depicted in Fig.~\ref{fig:intro}, affecting DNN performance and efficiency.
In the figure, each area with the same color represents elements sharing a common quantization scale factor: weights in (a) to (c) and partial-sums in (d) to (f).
\begin{table}[tbp]
    \centering
    \caption{Related Works on Partial-sum Quantization}
    \vspace*{-3mm}
    \label{tab:related works}
    \includegraphics[width=\columnwidth]{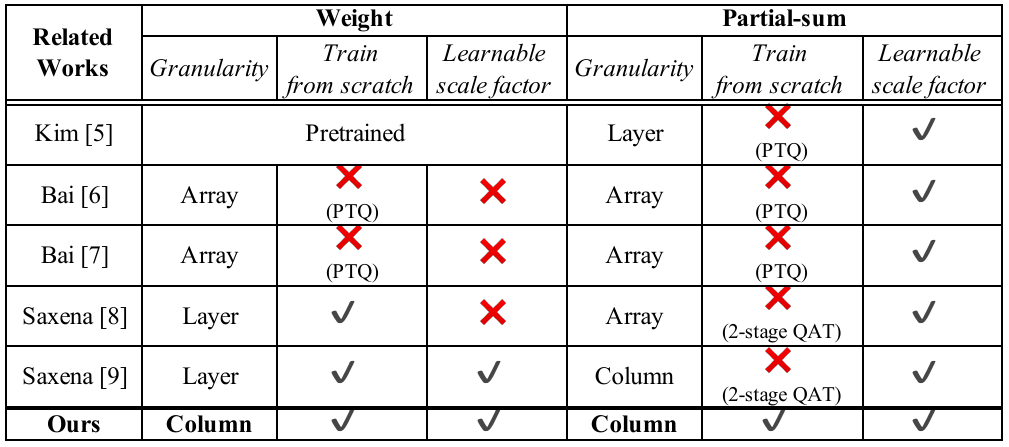}
    \vspace*{-10mm}
\end{table}

In \cite{b5}, the quantization scheme illustrated in Fig.~\ref{fig:intro}(a) and (d) is used, relying on post-training quantization (PTQ).
While PTQ simplifies the quantization process, it often results in critical quantization errors since the model is not trained for lower precision.
Furthermore, although a learnable scale factor is applied to partial-sums, the absence of a learnable scale factor for weights limits the model's ability to fully optimize its performance in a quantized setting.

The quantization approaches adopted by \cite{b6} and \cite{b7} correspond to the methods in Fig.~\ref{fig:intro}(b) for weights and (e) for partial-sums, but the adaptability is still constrained by PTQ.
The lack of learnable scale factors for weights constrains the model’s ability to adjust to quantization, making it less robust in handling quantization errors.

The authors in \cite{b8} and \cite{b9} adopted layer-wise granularity for weights as seen in Fig.~\ref{fig:intro}(a), but partial-sums are quantized differently—array-wise in \cite{b8} and column-wise in \cite{b9}.
Both methods rely on a two-stage QAT approach due to the mismatch between weight and partial-sum granularities, where weights undergo QAT from scratch, while partial-sums are only quantized during the second stage of training.
This leads to inefficiencies, as weights are overfitted to full-precision partial-sums during the first stage, delaying the model's adaptation to partial-sum quantization errors and hindering optimization.
Moreover, applying a learnable scale factor only to the partial-sums, as in \cite{b8}, restricts the model’s flexibility to adapt across both weights and partial-sums simultaneously, reducing its overall effectiveness.

In summary, the related works did not implement fine-grained quantization for both weights and partial-sums, nor did they apply learnable scale factors to both.
Additionally, these approaches left room for improving training efficiency.
Our approach addresses these limitations by aligning the quantization granularities of weights and partial-sums to column-wise level.
Table~\ref{tab:related works} compares recent works and our method across three key factors: the finest quantization granularity, the ability to train from scratch, and the use of learnable scale factors.

\section{Proposed Method}\label{sec:method}
In this section, we introduce our column-wise quantization approach for both weights and partial-sums, evaluating its significance in relation to dequantization overhead and training efficiency.
Additionally, we present our CIM-oriented convolution framework that seamlessly reflects the hardware architecture, addressing the inefficiencies associated with implementing column-wise quantization.

\subsection{Column-wise Weight and Partial-sum Quantization}

\begin{figure}[tbp]
    \centering
    \includegraphics[width=\columnwidth]{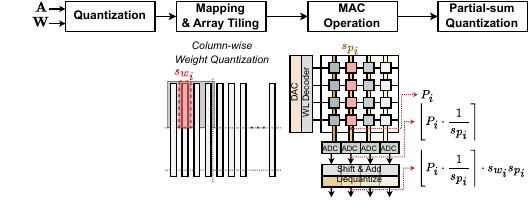}
    \vspace*{-8.5mm}
    \caption{
    Matrix multiplication of a DNN layer with the proposed column-wise weight and partial-sum quantization. ($A$: activations, $W$: weights, $P_i$: partial-sums, $s_{w_i}$: scaling factor for weights, $s_{p_i}$: scaling factors for partial-sums)
    }
    \label{fig:conv}
    \vspace*{-8mm}
\end{figure}

As illustrated in Fig.~\ref{fig:conv}, in typical CIM-based architectures, the convolution operation begins with the independent quantization of weights, $\mathbf{W}$, and activations, $\mathbf{A}$.
The quantized weights are mapped and tiled into arrays, which are then used in the MAC operation on each array, followed by the quantization of the resulting partial-sums.

Expanding on this workflow, our method improves it by employing column-wise quantization for both weights and partial-sums, as shown in Fig.~\ref{fig:conv}.
For instance, a column of weights in an array is quantized and multiplied by the corresponding quantized activation, generating the partial-sum as follows,
\begin{equation}
    P_i = \left\lfloor \frac{W_i}{s_{w_i}} \right\rceil A_{q_i},
\end{equation}
where $\left\lfloor z \right\rceil$ rounds $z$ to the nearest integer.
After the MAC operation, the resulting partial-sums are quantized:
\begin{equation}
    \left\lfloor \left\lfloor \frac{W_i}{s_{w_i}} \right\rceil A_{q_i} \cdot \frac{1}{s_{p_i}}\right\rceil=\left\lfloor P_i\cdot \frac{1}{s_{p_i}}\right\rceil.
\end{equation}

As demonstrated in the equations, the weight and partial-sum quantization processes are clearly distinct due to the separate rounding functions, enabling independent optimization for better flexibility and precision in the overall quantization process.
As the quantizer, we employ the LSQ\cite{b10} method to train scale factors for both weights and partial-sums separately, optimizing the model for fine-grained quantization effectively.
In addition, we extend LSQ to support scale factors at varying granularities, including column-wise quantization.

Finally, each partial-sum is dequantized: 
\begin{equation}
    \left\lfloor \left\lfloor \frac{W_i}{s_{w_i}} \right\rceil A_{q_i} \cdot \frac{1}{s_{p_i}}\right\rceil \cdot s_{w_i}s_{p_i}=\left\lfloor P_i\cdot \frac{1}{s_{p_i}}\right\rceil \cdot s_{w_i}s_{p_i}.
\end{equation}
If, on the other hand, weights are quantized at the layer-wise level, the weight quantization is simplified with a single scale factor, $s_w$, for all weights in the layer. 
However, when partial-sums remain quantized column-wise, the rest of the process for each $N\times N$ array follows the same steps as in the former case:
\begin{equation}
    \begin{split}
        &\left[ 
        \left\lfloor \left\lfloor \frac{W_1}{s_{w}} \right\rceil A_{q_1} \cdot \frac{1}{s_{p_1}}\right\rceil \cdot s_{w}s_{p_1}, \right.\\
        &... \: , \left\lfloor \left\lfloor \frac{W_N}{s_{w}} \right\rceil \left. A_{q_N} \cdot \frac{1}{s_{p_N}}\right\rceil \cdot s_{w}s_{p_N}\right].
    \end{split}
\end{equation}

To summarize, we propose column-wise quantization for both weights and partial-sums, which provides superior accuracy compared to other quantization schemes while maintaining the same dequantization overhead.
Also, the column-wise granularity alignment enhances training efficiency by enabling precise one-stage QAT, as further detailed in the following sections.
Moreover, it exhibits robustness against device-level variations, as the independent scale factors are assigned to each column effectively. 

\subsection{Dequantization in Column-wise Quantization}\label{sec:dequan}
\begin{figure}[tbp]
    \centering
    \includegraphics[width=\columnwidth]{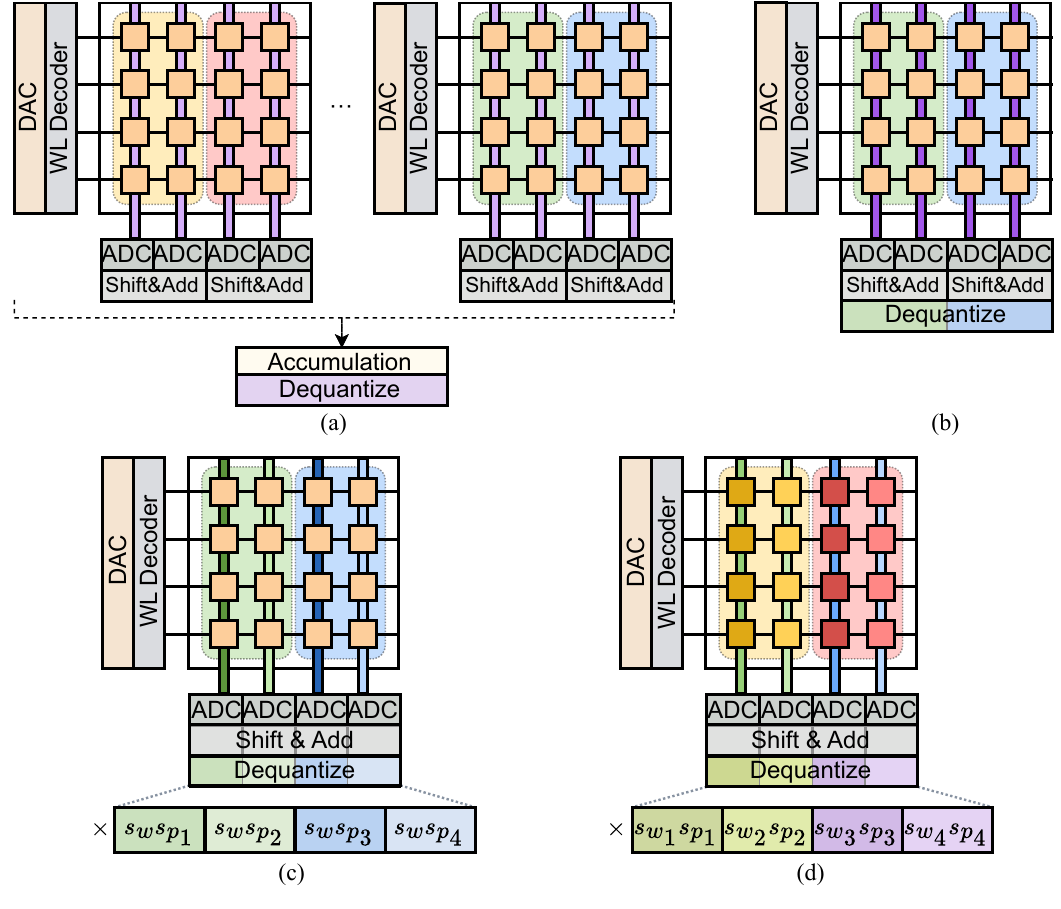}
    \vspace*{-10.5mm}
    \caption{
    Dequantization process after matrix multiplication in CIM.
    From (a) to (c), weights are quantized layer-wise, with partial-sums quantized (a) layer-wise, (b) array-wise, or (c) column-wise.
    (d) shows our proposed column-wise weight and partial-sum quantization.
    This figure assumes that the 4-bit weights are implemented with two 2-bit cells. 
    Memory cells with the same background color belong to the same output channel.
    }
    \label{fig:dequan arch}
    \vspace*{-6.5mm}
\end{figure}

Although distinct scale factors are required for weights and partial-sums as previously discussed, the hardware architecture allows for their dequantization to be performed simultaneously, thereby enhancing efficiency.
Fig.~\ref{fig:dequan arch} provides a visual representation of this process, with weights quantized layer-wise and partial-sums quantized (a) layer-wise, (b) array-wise, or (c) column-wise.

In the conventional layer-wise quantization shown in Fig.~\ref{fig:dequan arch}(a), outputs from each array are first accumulated, followed by a single dequantization step for the entire layer, requiring the overhead of one scale factor multiplication.
On the other hand, if partial-sums are quantized at array-wise level, those from the same output channel within an array are first shift-and-added.
Subsequently, each channel undergoes separate dequantization as illustrated in Fig.~\ref{fig:dequan arch}(b), increasing the overhead to $n_{\mathrm{array}}\times n_{\mathrm{oc}}$ multiplications for a layer, where $n_{\mathrm{array}}$ and $n_{\mathrm{oc}}$ are the number of arrays and output channels in an array.
Further granularity refinement is seen in Fig.~\ref{fig:dequan arch}(c), where each column of partial-sums has its unique scale factor, $s_{p_1}$ to $s_{p_4}$.
While the layer-wise weight scale factor, $s_w$, is shared across all columns, the system must store and apply the appropriate multiplied scale factor to each column, resulting in per-column overhead of $n_{\mathrm{split}}\times n_{\mathrm{array}}\times n_{\mathrm{oc}}$ multiplications, where $n_{\mathrm{split}}$ is the number of bit-splits.

Our quantization scheme extends the concept of column-wise quantization by applying unique scale factors to each column for both weights, $s_{w_1}$ to $s_{w_4}$, and partial-sums, $s_{p_1}$ to $s_{p_4}$.
The multiplied scale factors are applied to each partial-sum that flows out from its corresponding column, as shown in Fig.~\ref{fig:dequan arch}(d).
This approach appears to intensify the dequantization complexity due to the fine weight granularity.
However, the key discovery is that, without introducing additional dequantization overhead, aligning column-wise quantization for weights and partial-sums enables finer and more precise control over the quantization process, compared to the layer-wise weight with column-wise partial-sum configuration.

\subsection{A Convolution Framework for Column-wise Quantization}
\begin{figure}[tbp]
    \centering
    \includegraphics[width=\columnwidth]{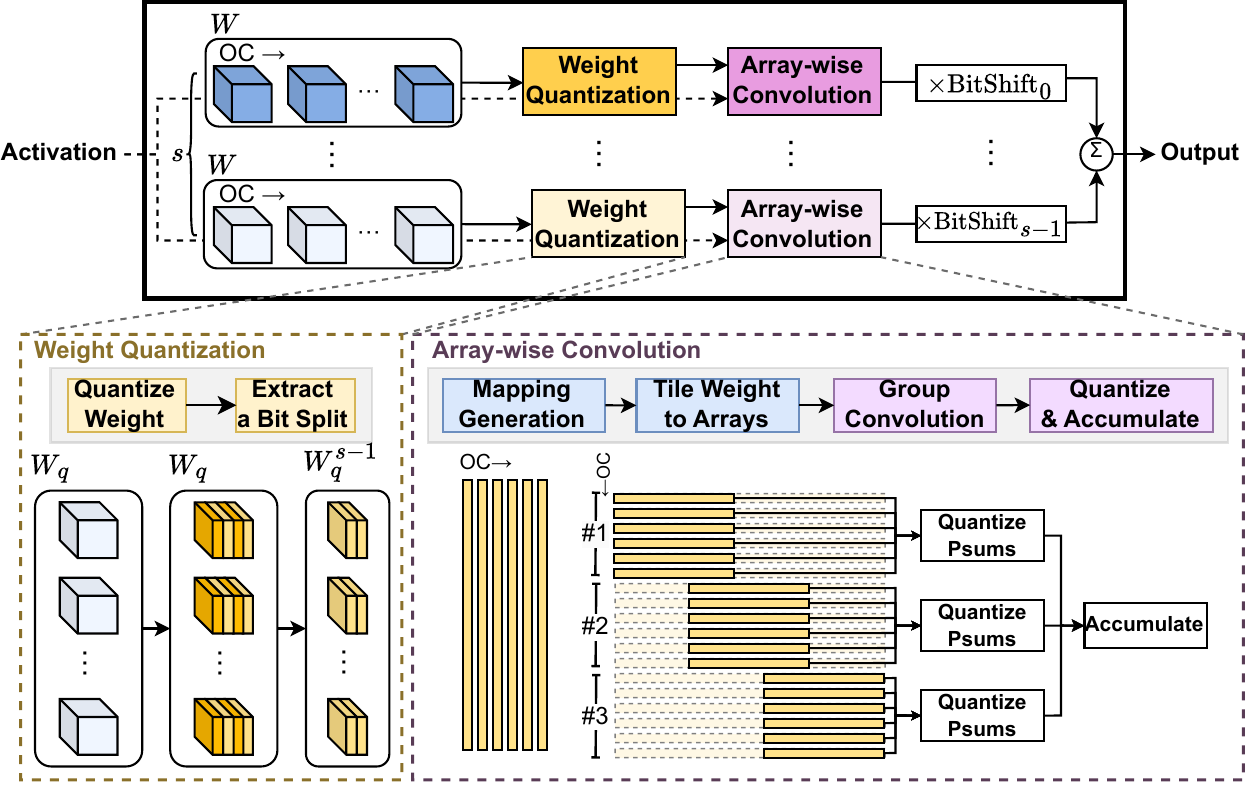}
    \vspace*{-8.5mm}
    \caption{Convolution framework overview for bit-scalable CIM.
    The figure outlines the process, starting with weight duplication and quantization into bit-splits.
    Each bit-split undergoes array-wise convolution with weight mapping, tiling, and partial-sum quantization.
    Finally, outputs from each split are bit-shifted and accumulated to form the final convolution output.}
    \label{fig:framework}
    \vspace*{-7.5mm}
\end{figure}

We propose a custom convolution layer framework designed to efficiently implement convolution layers with column-wise weight and partial-sum quantization.
Conventionally, handling individual columns introduces challenges of time costs and complexities in data management, often resulting in inefficiencies.
We resolve these obstacles by utilizing a unique array tiling method combined with group convolution.
As illustrated in Fig.~\ref{fig:framework}, we propose solutions to overcome these challenges and enhance processing efficiency, which emulates the convolution process in bit-scalable CIM architectures.

In CIM arrays, quantized weights break down into smaller segments, bit-split weights, to fit the number of capable bits per memory cell.
When replicating bit-scalable operations, it is necessary to enable access to each bit-split of the weight.
To facilitate this, we duplicate the original weight, $W$, according to the number of bit-splits, $s$, allowing independent processing of each bit-split during the quantization and convolution stages.

Subsequently, these duplicated weights are quantized to $W_q$ at layer-wise, array-wise, or column-wise granularity.
Each quantized weight then undergoes array-wise MAC operation, facilitated by weight mapping and tiling.
In the conventional im2col method, time-consuming linear operations are used to compute MAC results, creating a noticeable bottleneck.
To improve processing efficiency, we propose a novel tiling method that transforms the linear operation into a convolution.
By strategically adjusting the tiling stride, we ensure stretched kernels remain intact in each array, as described in Fig.~\ref{fig:framework}, and reshape it into a 4-dimensional convolutional weight.

Moreover, sequential array-wise convolution introduces critical overhead, as it requires indexing arrays one by one.
To mitigate this challenge, we utilize group convolution, matching the number of groups to the number of arrays, as illustrated in Fig.~\ref{fig:framework}.
This removes sequential indexing delays and simplifies access to array-wise partial-sums, resulting in faster convolution and subsequent partial-sum quantization.

Following the array-wise convolution, the generated partial-sums are quantized and accumulated.
Importantly, weight and partial-sum granularities are independently selected from layer-wise, array-wise, or column-wise quantization, enabling tailored optimization based on the specific requirements of the model and hardware configuration.
Finally, the convolution outputs from each bit-split weight are shifted and accumulated to produce one convolutional layer output.

\subsection{Efficient One-stage QAT via Granularity Alignment}\label{sec:2-stage}
In \cite{b9}, a two-stage QAT approach was employed to manage the granularity mismatch between weights and partial-sums, with partial-sums only quantized in the second stage of training to reduce training costs.
In contrast, our method achieves efficient one-stage QAT from scratch by aligning the granularity of both weights and partial-sums to the column-wise level.
This approach simplifies training by eliminating the need for separate stages to handle the partial-sum quantization, and ensures consistent optimization of both without compromising granularities.

\section{Experimental Results}\label{sec:experiment}

\subsection{Settings}
We evaluated the impact of quantization granularities on network accuracy and dequantization overhead, applying one-stage QAT of ResNet-20\cite{b3} on CIFAR-10 and CIFAR-100, and ResNet-18\cite{b3} on ImageNet\cite{b4}.
Each experiment employed distinct quantization and array size settings, as detailed in Table~\ref{tab:settings}.
For ResNet-20, we compared accuracy across layer-wise, array-wise, and column-wise quantization for weights and partial-sums, while, for ResNet-18, we replicated granularities from related works for direct comparison.
Additionally, we examined the effect of granularity alignment on the dequantization process and training efficiency, and conducted variation analysis based on the granularity combinations used in related works, applying ResNet-20 settings on CIFAR-10.

\begin{table}[tbp]
    \centering
    \caption{Experimental Settings for Evaluating\\ the Impact of Quantization Granularities on Accuracy}
    \vspace*{-3mm}
    \label{tab:settings}
    \includegraphics[width=\columnwidth]{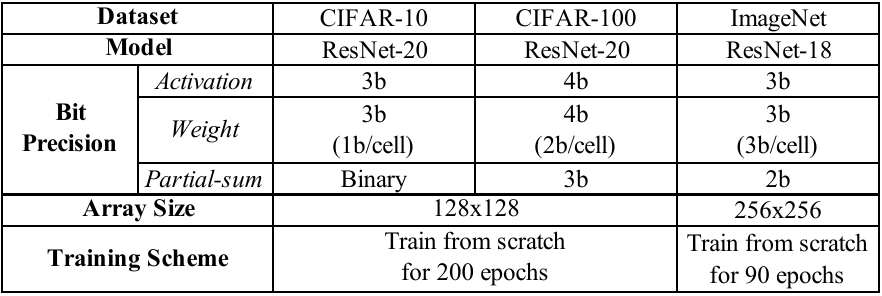}
    \vspace*{-9mm}
\end{table}

\subsection{Impact of Quantization Granularity}
\begin{figure}[tbp]
    \centering
    \includegraphics[width=\columnwidth]{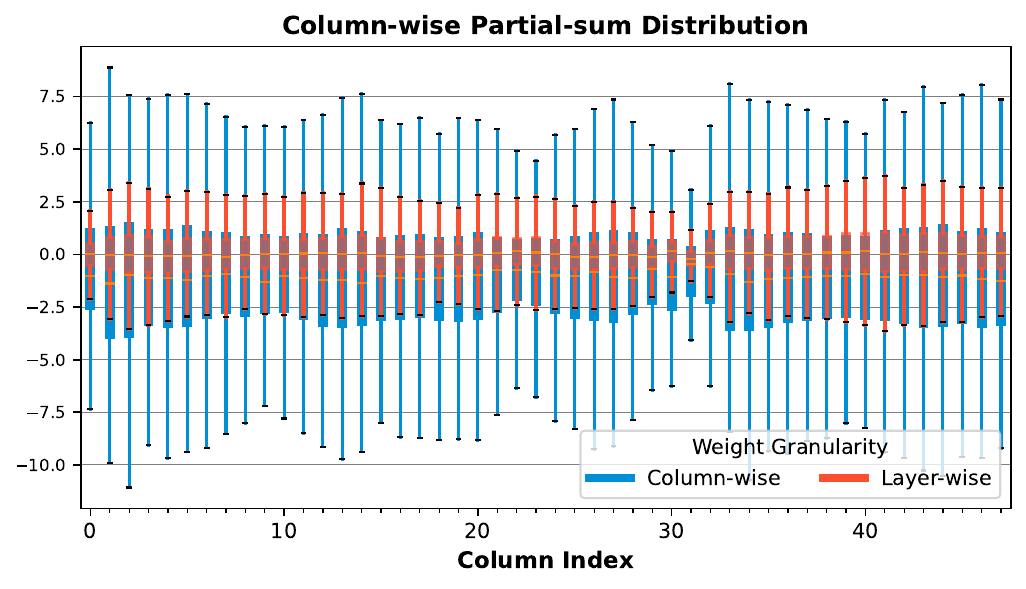}
    \vspace*{-8.5mm}
    \caption{
    Column-wise partial-sum distribution of 4th convolution layer of ResNet-20 for CIFAR-10, comparing dynamic ranges between layer-wise and column-wise weight quantization.
    }
    \label{fig:psum dstrb}
    \vspace*{-8mm}
\end{figure}

The granularity of weight quantization is essential for enhancing partial-sum representation capability as well as that of weights.
Fig.~\ref{fig:psum dstrb} compares the integer-valued column-wise partial-sum distributions for a specific layer from ResNet-20.

The figure clearly demonstrates that the column-wise weight quantization yields a larger dynamic range for the partial-sums, improving their representation ability.
In contrast, layer-wise weight quantization uses a single scale factor for the entire layer, resulting in a more uniform distribution that limits the adaptability to differences across columns.
By assigning distinct scale factors to each column, our column-wise weight quantization captures the weights more accurately, enabling more precise partial-sum computations.
Thus, this approach is particularly effective for fine-grained partial-sum quantization, surpassing coarser schemes.

\begin{figure}[tbp]
    \centering
    \includegraphics[width=\columnwidth]{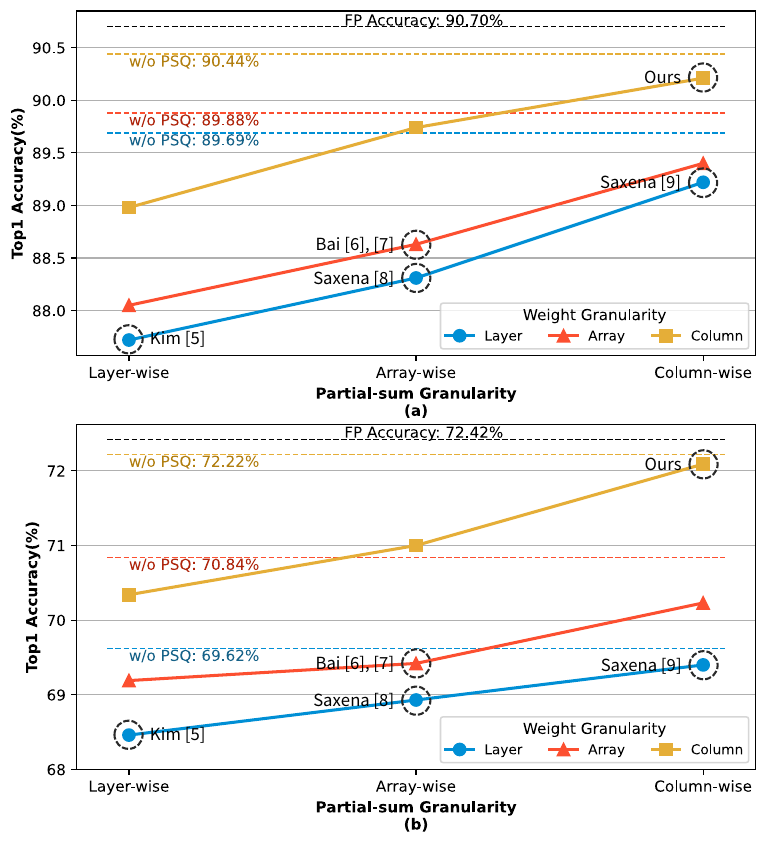}
    \vspace*{-9.5mm}
    \caption{
    Top1 inference accuracy of ResNet-20 on (a) CIFAR-10 and (b) CIFAR-100 with various granularity of weight and partial-sum quantization.
    Colored dashed lines represent the accuracy achieved without partial-sum quantization (PSQ) for each weight granularity.
    The accuracy results corresponding to the proposed method and previous works are highlighted.
    }
    \label{fig:cifar}
    \vspace*{-4.5mm}
\end{figure}

This advantage is further confirmed in our following experiments across ResNet20 on CIFAR-10 and CIFAR-100, and ResNet18 on ImageNet.
Fig.~\ref{fig:cifar}(a) presents a comparison of CIFAR-10 results with different granularity combinations.
The result presents that our proposed method attains the highest accuracy among all quantized models and related works.
Our method achieves a top-1 accuracy of 90.21\%, which is the closest to the full-precision(FP) model accuracy of 90.70\%, surpassing all related works.
Specifically, compared to \cite{b9}, which employed layer-wise weight and column-wise partial-sum quantization, our approach improves accuracy by 0.99\%.

This trend is observed in Fig.~\ref{fig:cifar}(b) for CIFAR-100 and Table~\ref{tab:imagenet} for ImageNet.

\begin{table}[tbp]
    \centering
    \caption{
    Inference accuracy of ResNet-18 on the ImageNet dataset
    }
    \vspace*{-3mm}
    \label{tab:imagenet}
    \includegraphics[width=\columnwidth]{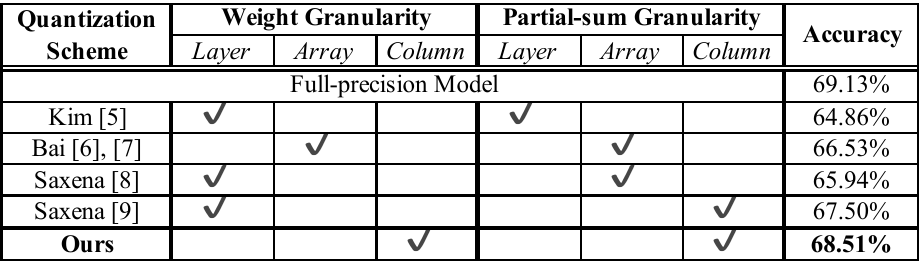}
    \vspace*{-10mm}
\end{table}
In each case, the model with column-wise quantization for both weights and partial-sums consistently outperforms those with coarser granularities.
For example, in CIFAR-100, column-wise weight and partial-sum quantization achieves a top-1 accuracy of 72.09\%, compared to 69.40\% achieved in \cite{b9} and 72.22\% in the baseline model without partial-sum quantization.
For ImageNet, our approach achieves 68.51\%, closest to the full-precision model accuracy at 69.13\%.

\subsection{Dequantization Analysis}\label{sec:dequan experiment}
\begin{figure}[tbp]
    \centering
    \includegraphics[width=\columnwidth]{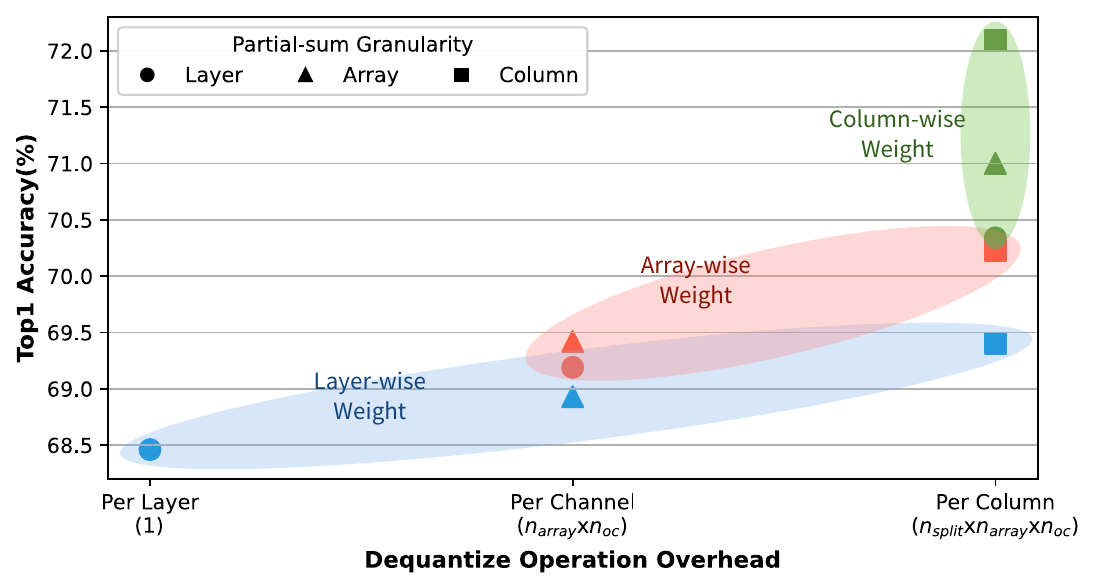}
    \vspace*{-9.5mm}
    \caption{
    Top-1 inference accuracy and dequantize operation overhead of ResNet-20 on CIFAR-100 with various granularities of weight and partial-sum quantization.
    Weights and activations are quantized to 4 bits, and 2-bit-per-cell arrays are used for the evaluation.
    }
    \label{fig:acc-overhead}
    \vspace*{-5mm}
\end{figure}

In Fig.~\ref{fig:acc-overhead}, all possible quantization schemes are categorized based on dequantize operation overhead per layer along the x-axis.
The results confirm that using finer weight quantization granularity achieves higher accuracy under the same dequantization overhead.
This shows the advantage of column-wise weight granularity in improving model performance without increasing hardware complexity.

\subsection{Impact of One-stage Quantization-aware Training}
\begin{figure}[tbp]
    \centering
    \includegraphics[width=\columnwidth]{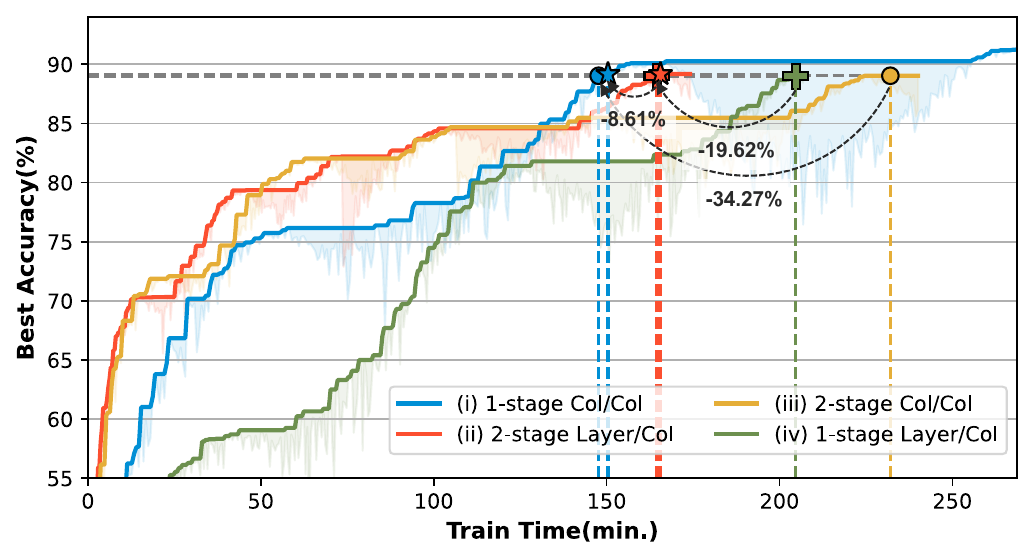}
    \vspace*{-8.5mm}
    \caption{
    Comparison of QAT schemes in terms of accuracy and train time.
    `A/B' represents the granularity of `weight/partial-sum' quantization.
    Plus marks show when case (ii) and (iv) achieved case (iv)'s best accuracy, circle marks indicate when case (i) and (iii) attained case (iii)'s best accuracy, and star marks denote when case (ii)'s best accuracy was reached in case (i) and (ii).}
    \label{fig:QATscheme}
    \vspace*{-5mm}
\end{figure}

Fig.~\ref{fig:QATscheme} compares four QAT schemes, each using different combinations of weight and partial-sum granularities.
The plus marks show that using the quantization scheme from \cite{b9}, the two-stage QAT approach achieves comparable accuracy to one-stage QAT  with 19.62\% less training cost.
In contrast, the circle marks show that one-stage QAT yields a higher accuracy in our quantization scheme, and 34.27\% less training cost compared to its two-stage counterpart.

The results verify the impact of aligning weight and partial-sum quantization granularities.
While the weight granularity is coarser than that of the partial-sums in cases (ii) and (iv), both share column-wise granularity in cases (i) and (iii).
The training delay between weights and partial-sums in case (iii) causes the weights to become overly tuned to full-precision partial-sums during the first stage, hindering performance in the second stage.
Moreover, the star marks indicate that case (i) achieves the highest accuracy of the case (ii) with 8.61\% less training cost, further validating our approach that the granularity alignment at the column-wise level ensures a more straightforward and efficient training process.

\subsection{Evaluation of Variation Robustness}
Non-idealities in nonvolatile memory, such as device variations, cause accuracy degradation in CIM accelerators.
In this section, we conducted a variation analysis on models using our proposed quantization scheme and those of related works.
As described in \cite{b11}, memory device variations are modeled by a log-normal distribution with a mean of zero.
To assess robustness, we introduced log-normal noise to the weights as follows,
\begin{equation}
    w_{var}=w\cdot e^{\theta},
\end{equation}
where $\theta$ is the noise, which follows a normal distribution with a zero mean, $w$ is the ideal weight, and $w_{var}$ is the weight after variation.

\begin{figure}[tbp]
    \centering
    \includegraphics[width=\columnwidth]{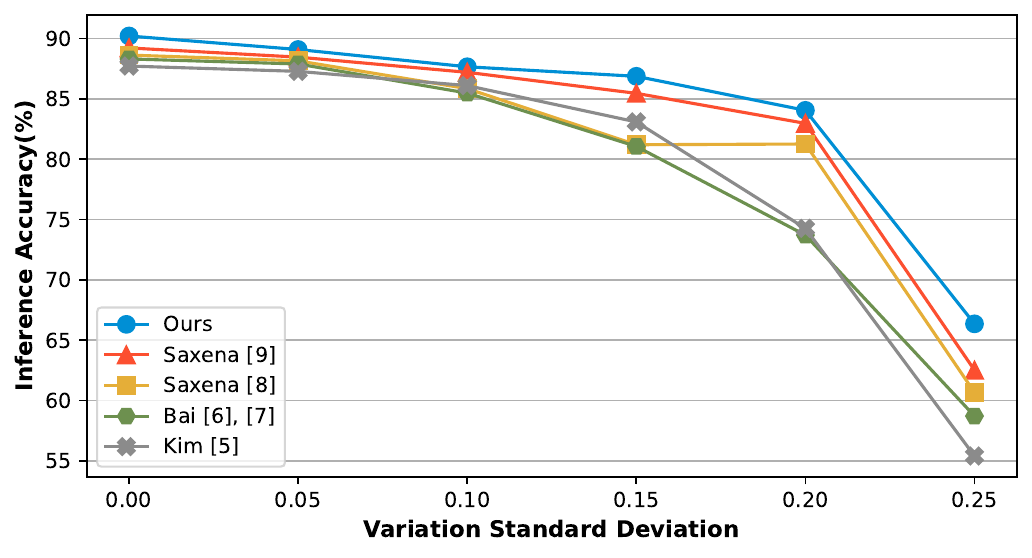}
    \vspace*{-9mm}
    \caption{Inference accuracy across different standard deviations of memory cell variation, comparing our quantization scheme with those of related works.}
    \label{fig:variation}
    \vspace*{-6mm}
\end{figure}

We evaluated inference accuracy across various standard deviations, as illustrated in Fig.~\ref{fig:variation}.
The result shows that models trained with our column-wise quantization method consistently achieve higher inference accuracy across all levels of variation, outperforming other quantization schemes.
This gap highlights the robustness of our column-wise quantization approach in preserving model performance under hardware-induced variations, delivering both higher accuracy and greater resilience to memory cell variations.

\section{Conclusion}
We propose an innovative quantization strategy that aligns weight granularity with partial-sums at the column-wise level.
Our method improves accuracy without increasing dequantization overhead and enhances training efficiency by removing the need for two-stage training.
Although managing fine-grained weights and partial-sums presents challenges, we address them through an open-source CIM-oriented convolution framework, incorporating a novel array tiling method and group convolution.
Our experiments on ResNet-20 (CIFAR-10 and CIFAR-100) and ResNet-18 (ImageNet) show substantial accuracy improvements—0.99\%, 2.69\%, and 1.01\%, respectively—over leading methods.
Additionally, the method's robustness to memory cell variations confirms its effectiveness in enhancing both accuracy and hardware efficiency in CIM-based accelerators.

\section*{Acknowledgement}
This work was partly supported by
the National Research Foundation of Korea (NRF) grant (No. RS-2024-00345732), 
the Institute for Information \& communications Technology Planning \& Evaluation (IITP) grants (RS-2020-II201821, IITP-2021-0-02052, RS-2019-II190421, RS-2021-II212068), 
the Technology Innovation Program (RS-2023-00235718, 23040-15FC) funded by the Ministry of Trade, Industry \& Energy (MOTIE, Korea) (1415187505), 
and Samsung Electronics Co., Ltd (IO230404-05747-01). 


\end{document}